\documentclass[%
 reprint,
showpacs,
showkeys,
 amsmath,amssymb,
 prl,
]{revtex4-1}

\pdfoutput=1
\usepackage{graphicx}
\usepackage{dcolumn}
\usepackage{bm}
\usepackage{wrapfig}
\usepackage{amssymb}
\usepackage{epstopdf}
\usepackage{float}
\usepackage{enumitem}
\setlist[enumerate]{itemsep=0mm}

\usepackage{lipsum}
\makeatletter
\newcommand*{\balancecolsandclearpage}{%
  \close@column@grid
  \cleardoublepage
  \twocolumngrid
}
\makeatother

\begin{document}


\title{Graded, Dynamically Routable Information Processing\\ with Synfire-Gated Synfire Chains}

\author{Zhuo Wang}
\affiliation{Center for Bioinformatics, National Laboratory of Protein Engineering and Plant Genetic Engineering, College of Life Sciences, Peking University, Beijing, China\\ 
wangz@mail.cbi.pku.edu.cn}

\author{Andrew T. Sornborger}
\affiliation{Department of Mathematics, University of California, Davis, USA\\
ats@math.ucdavis.edu}

\author{Louis Tao}
\affiliation{Center for Bioinformatics, National Laboratory of Protein Engineering and Plant Genetic Engineering, College of Life Sciences, and Center for Quantitative Biology, Peking University, Beijing, China\\
taolt@mail.cbi.pku.edu.cn\\
\\
{\rm ATS and LT are corresponding authors.}\\}

\date{\today}

\maketitle

\onecolumngrid

\section{Abstract}
\noindent
Coherent neural spiking and local field potentials are believed to be signatures of the binding and transfer of information in the brain. Coherent activity has now been measured experimentally in many regions of mammalian cortex. Synfire chains are one of the main theoretical constructs that have been appealed to to describe coherent spiking phenomena. However, for some time, it has been known that synchronous activity in feedforward networks asymptotically either approaches an attractor with fixed waveform and amplitude, or fails to propagate. This has limited their ability to explain graded neuronal responses. Recently, we have shown that pulse-gated synfire chains are capable of propagating graded information coded in mean population current or firing rate amplitudes. In particular, we showed that it is possible to use one synfire chain to provide gating pulses and a second, pulse-gated synfire chain to propagate graded information. We called these circuits synfire-gated synfire chains (SGSCs). Here, we present SGSCs in which graded information can rapidly cascade through a neural circuit, and show a correspondence between this type of transfer and a mean-field model in which gating pulses overlap in time. We show that SGSCs are robust in the presence of variability in population size, pulse timing and synaptic strength. Finally, we demonstrate the computational capabilities of SGSC-based information coding by implementing a self-contained, spike-based, modular neural circuit that is triggered by, then reads in streaming input, processes the input, then makes a decision based on the processed information and shuts itself down.
\vskip .6cm

\twocolumngrid

%

\section{Introduction}

Accumulating experimental evidence implicates coherent activity as an important element of cognition. Since its discovery \citep{GrayEtAl1989}, activity in the gamma band has been demonstrated to exist in numerous regions of the brain, including hippocampus \citep{BraginEtAl1995,CsicsvariEtAl2003,ColginEtAl2009}, 
numerous areas in cortex \citep{GrayEtAl1989,Livingstone1996,WomelsdorfEtAl2007,BroschEtAl2002,BauerEtAl2006,PesaranEtAl2002,BuschmanMiller2007,MedendorpEtAl2007,BuschmanMiller2007,GregorgiouEtAl2009,SohalEtAl2009}, 
amygdala and striatum \citep{PopescuEtAl2009}. Gamma band activity is associated with sharpened orientation \citep{AzouzGray2000} and contrast \citep{HenrieShapley2005} tuning in V1, and speed and direction tuning in MT \citep{LiuNewsome2006}. Attention has been shown to increase gamma synchronization between V4 and FEF \citep{GregorgiouEtAl2009}, LIP and FEF \citep{BuschmanMiller2007}, V1 and V4 \citep{BosmanEtAl2012}, and MT and LIP \citep{SaalmannEtAl2007}; In general, communication between sender and receiver neurons is improved when consistent gamma-phase relationships exist between upstream and downstream sites \citep{WomelsdorfEtAl2007}.

Theta-band oscillations are associated with spatial memory \citep{OKeefe1993,Buzsaki2002}, where neurons encoding the locations of visual objects and an animal's own position have been identified \citep{OKeefe1993,SkaggsEtAl1996}. Loss of theta results in spatial memory deficits \citep{Winson1978} and pharmacologically enhanced theta improves learning and memory \citep{MarkowskaEtAl1995}.

Classical coding mechanisms are related to neural firing rate \citep{AdrianZotterman1926}, population activity \citep{HubelWiesel1965,HubelWiesel1968,KaisslingPriesner1970}, and spike timing \citep{pmid8768391}. Firing rate \citep{AdrianZotterman1926} and population codes \citep{Knight1972,Knight2000,SirovichEtAl1999,Gerstner1995,BrunelHakim1999} are two ways to average spike number to represent graded information, with population codes capable of faster  processing since they average across many fast responding neurons. Thus population and temporal codes can make use of the sometimes millisecond accuracy \citep{pmid8768391,pmid17805296,pmid23010933} of spike timing to represent dynamic averages.

New mechanisms have been proposed for short-term memory \citep{LismanIdiart1995,JensenLisman2005,Goldman2008}, information transfer via spike coincidence \citep{Abeles1982,KonigEtAl1996,Fries2005} and information gating \citep{SalinasSejnowski2001,Fries2005,RubinTerman2004,pmid24730779,pmid25503492} that rely on coherent activity in the gamma- and theta-band. For example, Fries's communication-through-coherence (CTC) model \citep{Fries2005} makes use of synchronous input that can provide windows in time during which spikes may be more readily transferred across a synapse. Additionally, synchronous firing has been used in Abeles's synfire network \citep{Abeles1982,KonigEtAl1996,pmid10591212,pmid21106815,KistlerGerstner2002} giving rise to volleys of propagating spikes.

Research to understand the network mechanisms linking coherent activity and information transfer has largely focused on synfire chains \cite{pmid12684488,pmid10591212,pmid12730700,KistlerGerstner2002,pmid24298251,pmid16641232}. The hope has been that synfire chains could be used to understand rapid, feedforward computation across multiple regions of cortex, as is seen in the response to rapid visual categorization experiments \citep{pmid11253215}.  However, many studies have shown that although it is possible to transfer volleys of spikes stably from layer to layer, the spike probability waveform tends to an attractor with fixed amplitude \cite{pmid12684488,pmid10591212,KistlerGerstner2002}. Below, we refer to these chains as ``attractor synfire chains". In attractor synfire chains, although a volley of spikes can propagate, graded information, in the form of a rate amplitude, cannot.

Additional numerical studies investigating information propagation have shown that it is possible to transfer firing rates through feed-forward networks when there is sufficient background activity to keep the network near threshold \cite{pmid11880526}. This mechanism has the disadvantage that firing rate information cannot be controlled other than by increasing or decreasing background activity. Other studies have shown that additional coherent spatio-temporal structures ({\it e.g.} hubs or oscillations) can stabilize the propagation of synchronous activity and select specific pathways for signal transmission \cite{pmid24730779,pmid25503492,AkamKullmann2010,pmid24434912}. 

Recently, we showed that information contained in the amplitude of a synaptic current or firing rate may be faithfully ({\it exactly} in a mean-field model) transferred from one neuronal population to another \cite{SornborgerWangTao}. In that work, coherent, non-overlapping gating pulses provided a sequence of temporal windows during which graded information was successively integrated then transferred on the synaptic time scale. We further showed that a self-contained, feed-forward network in the synfire regime can propagate graded information. This circuit, the synfire-gated synfire chain (SGSC), used an attractor synfire chain to generate gating pulses for graded information transfer. The SGSC mechanism provides a neuronal population based means of dynamically routing {\it graded} information through a neural circuit. 

The SGSC mechanism naturally provides a framework in which information control and processing are separated into two complementary parts of a single neural circuit. Information processing is performed by synaptic connectivity as information propagates from layer to layer. Information control is performed by gating pulses that dynamically route information through the circuit \citep{SornborgerWangTao}.

Here, we show that, in general, overlapping gating pulses can be used to propagate temporally overlapping graded information. These solutions improve on our previous work by allowing information to cascade through a multi-layer network on a time scale that is a fraction of the synaptic time scale. We then describe a general class of time-translationally-invariant solutions to a mean-field model motivated by simulational results from the SGSC. We show a correspondence between the mean-field model and the SGSC and investigate robustness of the SGSC to finite-size effects, variability in the synaptic coupling and variability in the delay between pulses from layer to layer. 

Finally, we show that by combining graded information with gating pulses, conditional decisions may be made to control information flow and the subsequent processing performed by the circuit. In order to demonstrate this information coding and decision making framework, we implement a self-contained, spike-based, modular neural circuit that is triggered by an input stream, reads in and processes the input, generates a conditional output based on the processed information, then shuts itself off.

\section{Results}

\begin{figure*}[t]
  \includegraphics[width=\textwidth]{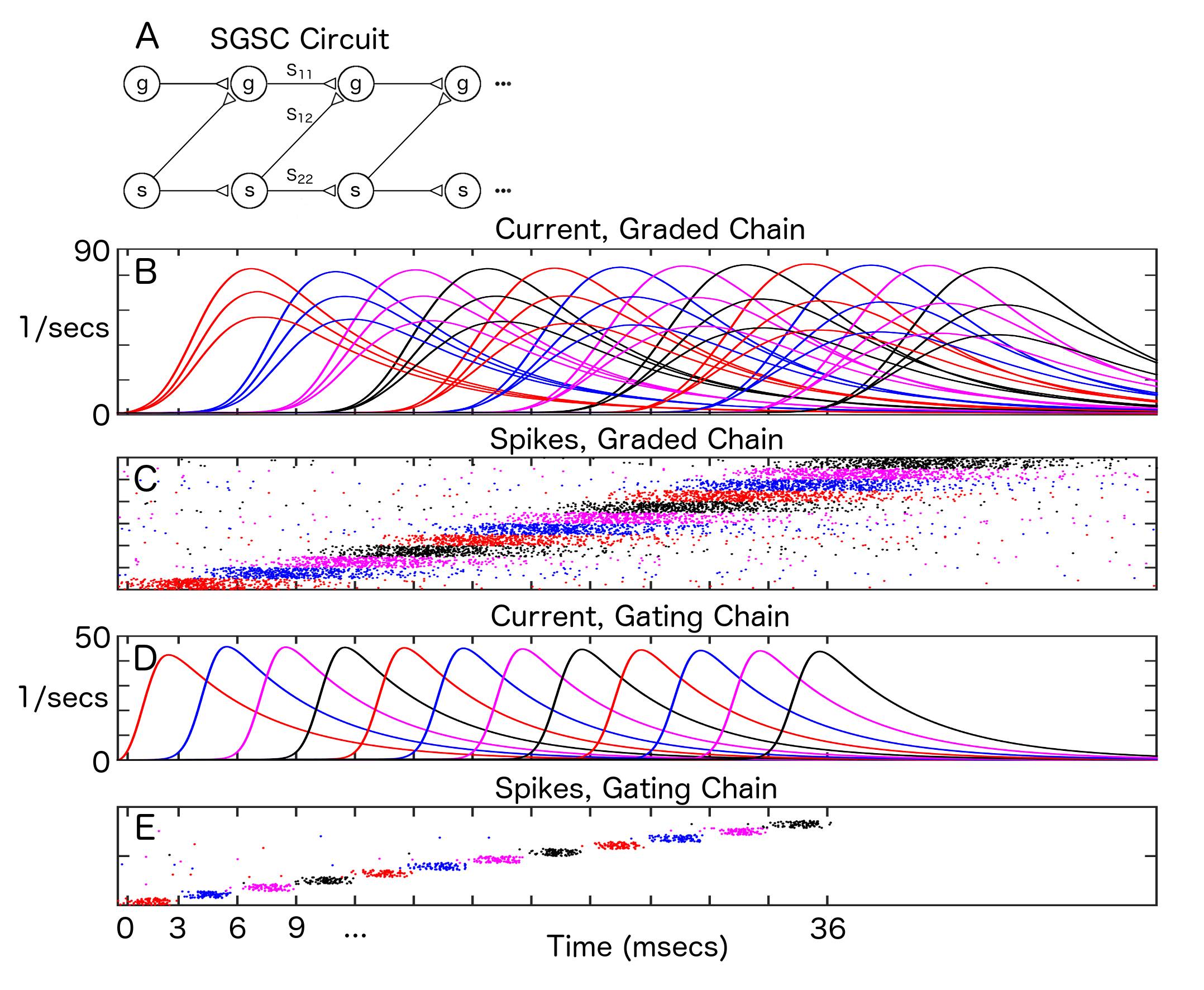}
  \caption{Graded information transfer in synfire-gated synfire chains. A) Circuit diagram. `g' denotes a population in the graded chain. `s' denotes a population in the gating chain. $S_{11}$, $S_{12}$ and $S_{22}$ denote synaptic couplings between and within the respective chains. The gating chain generates pulses that gate the propagation of graded information in the graded chain. B) Mean, synaptic current amplitude transferred across $12$ neural populations. N = 1000. Averaged over $50$ trials. Three amplitudes are depicted. C) Spike rasters from graded populations for one instance of graded transfer. D)  Mean, synaptic current amplitude for fixed amplitude synfire chain. N = 100. Averaged over $50$ trials. E) Spike rasters from gating populations for one instance of graded transfer.}
\end{figure*}

In Fig. 1, we show how graded information may be propagated in an SGSC neural circuit (Fig. 1A). This circuit consists of two feedforward networks. One network (gating chain), set up to operate in the attractor synfire regime, generates a fixed amplitude pulse that propagates from layer to layer (Fig. 1D,E). The second network (graded chain) receives gating pulses from the gating chain and is capable of propagating graded currents and firing rates from layer to layer (Fig. 1B,C). The gating chain delivers pulses offset by time $T_0$ to the graded chain rapidly enough that there is an overlap in the integration of graded information and its transmission from one layer to the next. Graded information, in the form of synaptic currents and firing rates, is faithfully propagated across all $12$ layers in the simulation.

\begin{figure}[t]
  \includegraphics[width=0.45\textwidth]{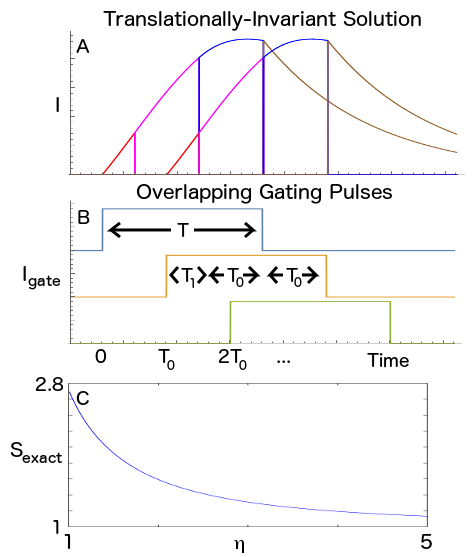}
    \caption{Graded information transfer with overlapping pulses, exact mean-field solution. A) Graded, mean current amplitudes across $2$ populations. Two overlapping solutions are shown, one upstream (earlier in time) and one downstream (later in time). The downstream current evolution is easiest to understand: The red segment depicts the epoch when the second gating pulse has brought the downstream population to threshold. During this time, the upstream current (depicted in magenta) is integrated and the downstream current begins to rise. Once the upstream current enters the next epoch (depicted in blue), the downstream current (depicted in magenta) continues to rise. After the upstream current begins to decay exponentially (depicted in brown), the downstream current continues to rise (depicted in blue) until the gating pulse ends. At this point, the downstream current decays exponentially. So, from the point of view of the downstream population, the red segment represents the integration of the pink segment of the upstream population, the magenta segment represents the integration of the blue segment of the upstream population, and the blue segment represents the integration of the brown segment of the upstream population. $T_0/\tau = 0.6$, $T_1/\tau = 0.3$, and $T/\tau = 1.5$. $S_{exact}$ for these values is $1.582$. The coefficients of the solution polynomial are $\{ 0.733, 0.640, 0.228 \}$ (See Appendices). B) Gating pulses offset from $0$ for clarity. C) $S_{exact}$ vs. $\eta$.}
\end{figure}

The observation that spike volleys in successive layers of the SGSC overlap in time led us to consider an extension of our previous mean-field model \cite{SornborgerWangTao} in which the integration of graded information in successive populations also overlapped in time. As in our previous work, we consider the idealized case in which the gating pulses are square. In Fig. 2, we show a translationally invariant solution (Fig. 2A) and gating pulses (Fig. 2B) from such a mean-field model. Successive gating pulses of length $T$ are offset by time $T_0$. The solution is divided into segments which are the result of the integration of spikes in the corresponding segment (shifted by $T_0$) from the previous layer during the gating pulse.

For fixed $T$ and $T_0$, we find time translationally-invariant solutions for special values of the feedforward coupling strength, $S = S_{exact}$ (see Materials and Methods and Appendices). In Fig. 2C, we plot $S_{exact}$ as a function of $\eta = T/T_0$, where $\eta$ is a measure of the overlap in the integration and transmission of graded information. Note that $S_{exact}$ becomes flatter as the overlap, $\eta$, gets larger. This implies that, for large overlaps, any propagation error in the solution due to deviations from $S_{exact}$ is small. Thus, in the large $\eta$ regime, information propagation is robust to variability in both pulse timing and coupling strength. For practical purposes, we find that $\eta > 2$ or $3$ is sufficiently robust. Furthermore, for a generic feedforward network, there exists a wide range of $S$ (roughly, $S$ from $1$ to $2.7$) where we can find time translationally-invariant solutions for which graded propagation is possible.

\begin{figure}[t]
  \includegraphics[width=0.45\textwidth]{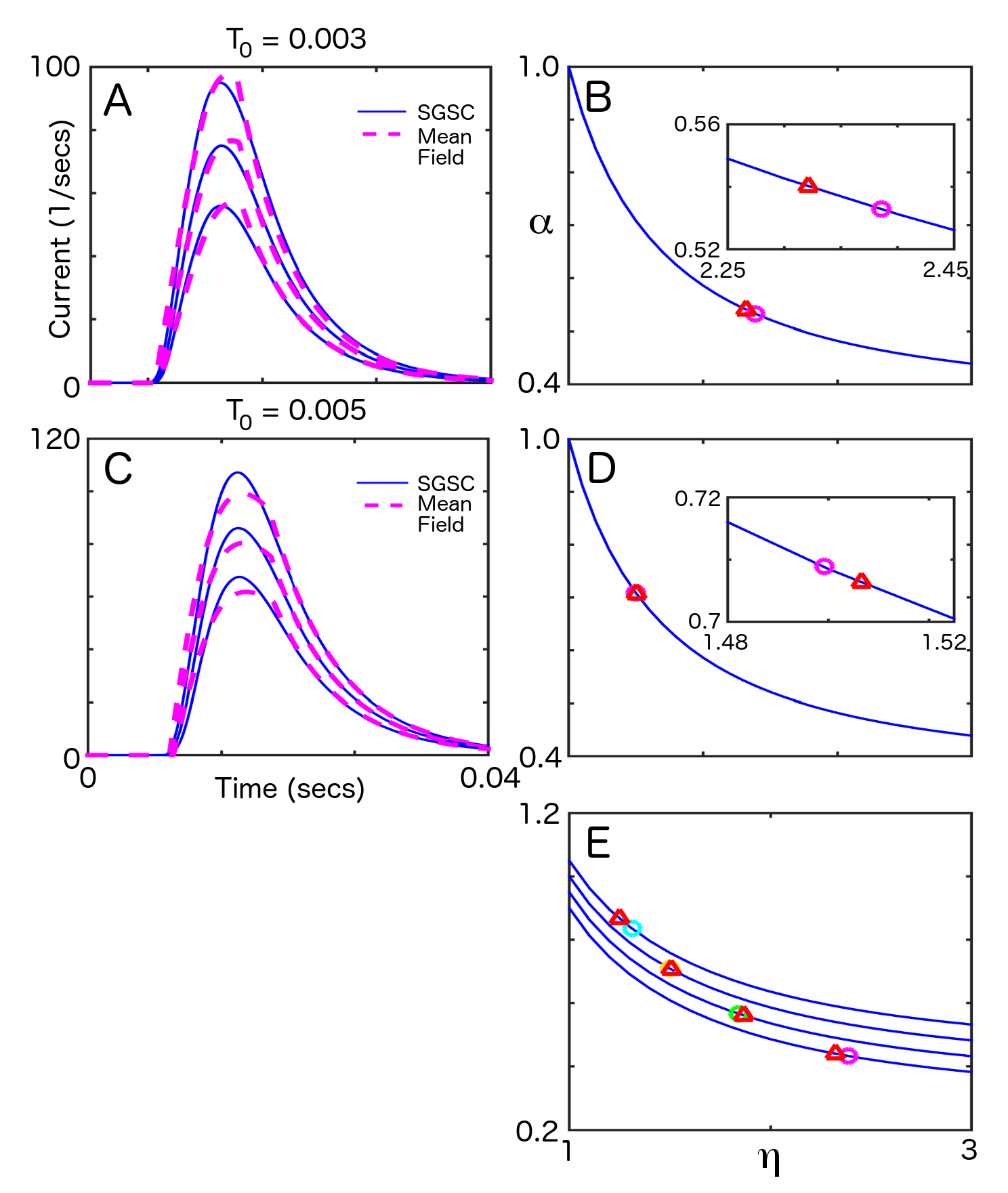}
   \caption{Fitting a square-pulse gated mean-field model of the SGSC. A) Fits of mean-field model and I \& F simulations for $3$ amplitudes for $T_0 = 0.003$. B) $T_0 = 0.003$: Blue line - $\alpha_{exact}$ as a function of $\eta$. Red triangle - $( \eta_{sim}, \alpha_{sim} )$, purple circle - $( \eta_{fit}, \alpha_{fit} )$. Inset: magnification showing location of results from fit. C) Fits of mean-field model and I \& F simulations for $3$ amplitudes for $T_0 = 0.005$. D) $T_0 = 0.005$: Blue line - $\alpha_{exact}$ as a function of $\eta$. Red triangle - $( \eta_{sim}, \alpha_{sim} )$, purple circle - $( \eta_{fit}, \alpha_{fit} )$. Inset: magnification showing location of results from fit. E) Results of fit for $T_0 = 0.003, 0.004, 0.005, 0.006$. Traces offset for clarity.}
\end{figure}

In Fig. 3, we explore whether our mean-field theory could be used to model our I\&F simulation results. First, we determined the parameters $(\eta_{fit}, \alpha_{fit})$ that gave the best-fitting mean-field solution to the simulation data, given known $T_0$. Here, we define $\alpha \equiv S (T_0/\tau) e^{-T_0/\tau}$, so that $\alpha = 1$ at $\eta = 1$. Next, using the simulational synaptic coupling, $\alpha_{sim}$, we found $\eta = \eta_{sim}$ that corresponded to the time-translationally invariant solution of the mean-field model. Closeness of these two points would give evidence that the mean-field theory, despite the simplications used to derive it ({\it e.g.} precisely timed square gating pulses, linear f-I curve, etc.), can be used to model the I\&F simulation. We show details of this fitting procedure for two different $T_0$'s (Fig. 3A--D), and summarize the results for $T_0 = 0.003, 0.004, 0.005, 0.006$ (Fig. 3E). The closeness of model fits with simulation results, for a wide range of overlaps, indicates that the mean-field theory is a good model of the SGSC simulation.

\begin{figure}[t]
  \includegraphics[width=0.45\textwidth]{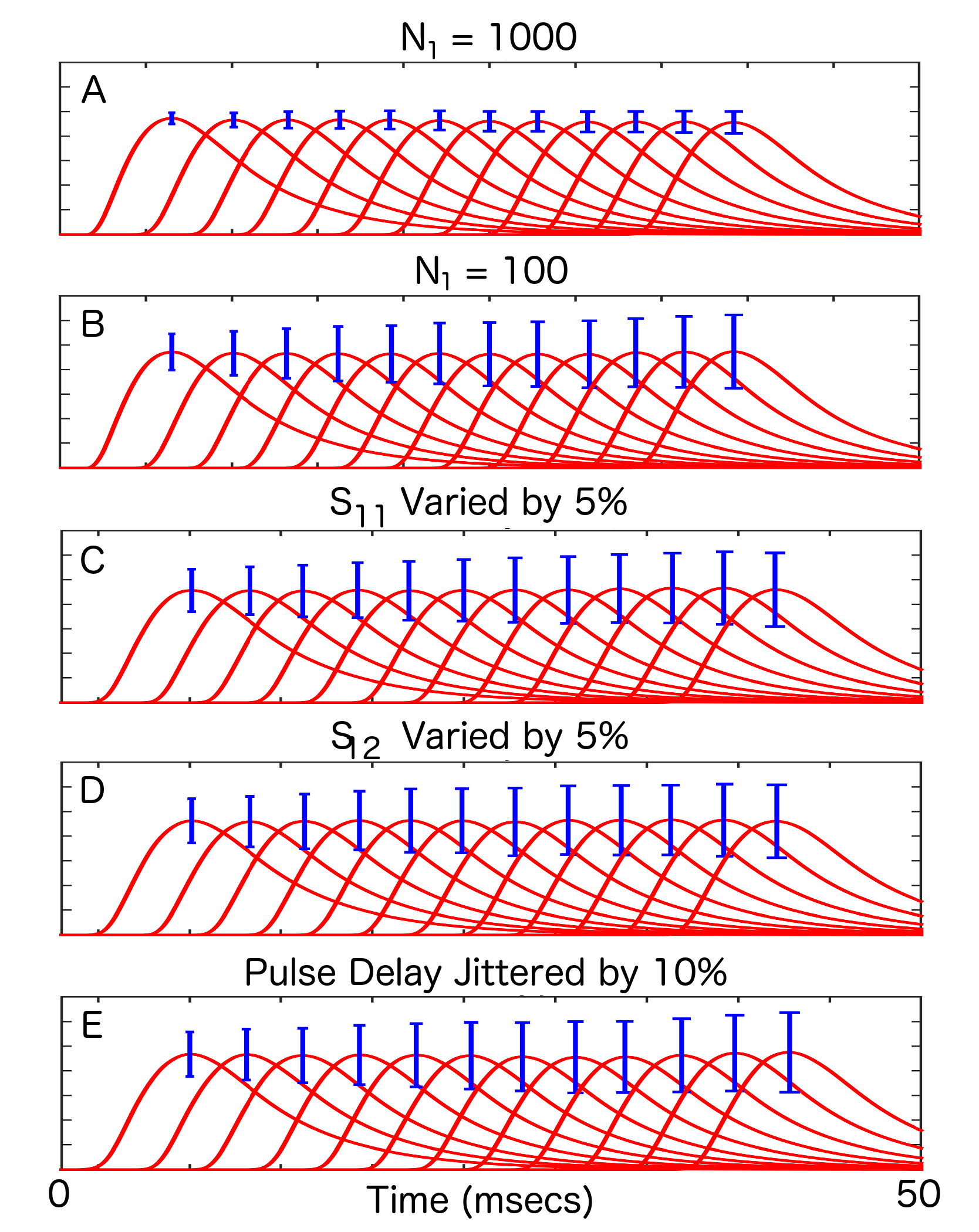}
  \caption{Signal-to-Noise-Ratio as a function of the number of transfers. Red - mean current amplitude for transfer across $12$ layers. Blue - standard deviation of current amplitude. Mean and standard deviation calculated from $1000$ trials. A) $N = 1000$, B) $N = 100$, C) $S_{11}$ taken from a uniform distribution with half-width of $5\%$, $N = 100$, D) $S_{12}$ taken from a uniform distribution with half-width of $5\%$, $N = 100$, E) Pulse delay jittered by $10\%$, $N = 100$.}
\end{figure}

In Fig. 4, we investigate the robustness of pulse-gated synaptic current transfer in the SGSC to finite-size effects, variability in synaptic coupling, and inaccuracies in pulse timing. As would be expected, transfer variability decreases as $1/\sqrt{N}$ (Fig. 4A,B). Randomness in synaptic coupling either in the gating chain or the coupling between chains has little effect on the variability (compare Fig. 4C,D with Fig. 4B). As we mentioned above, this is expected due to the flatness of $S_{exact}(\eta)$ for large $\eta$. Here, $\eta = 2.5$. Similarly, jittering $T_0$ has little effect on the variability of current transfer (Fig. 4E).

Pulse-gated propagation mechanisms, such as the SGSC, naturally give rise to a probabilistic, spike-based information processing framework in which information is processed by graded chains and the flow of information is controlled by gating chains \cite{SornborgerWangTao}. Additionally, logic operations may be performed by allowing graded information to interact with the pulse generator (see Materials and Methods).

\begin{figure*}[t]
  \includegraphics[width=\textwidth]{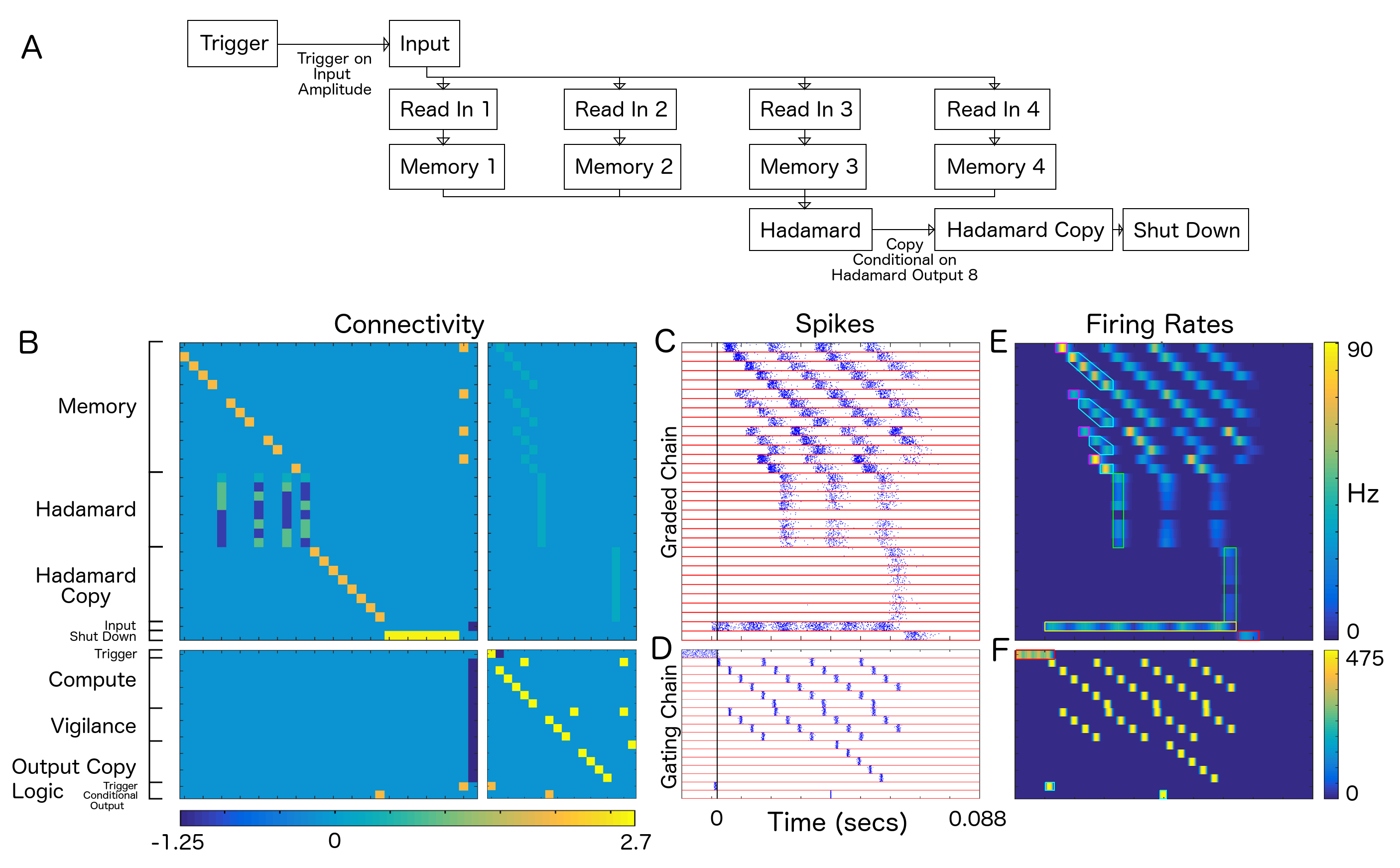}
  \caption{Autonomous decision making circuit. A) Neural Circuit. B) Connectivity matrix (four components) $K^{11}$ (graded to graded, upper left), $K^{12}$ (gating to graded, upper right), $K^{21}$ (graded to gating, lower left), and $K^{22}$ (gating to gating, lower right). Color bar denotes connectivities. Graded chain populations:  ``Memory" ($1$ - $14$), ``Hadamard" ($15$ - $22$), ``Hadamard Copy" ($23$ - $30$), ``Input" ($31$), and ``Shutdown" ($32$). Gating chain populations are: ``Trigger" ($33$), a re-entrantly coupled population that fires until inhibited. ``Compute" ($34$ - $39$) for gating the computation of the windowed Hadamard transform in the Memory and Hadamard populations, ``Vigilance" ($40$ - $43$) a pulse loop that, along with the ``Logic - Conditional Output" ($50$) population makes a decision based on the amplitude of the output of the $8$'th Hadamard population, and ``Output Copy" ($44$ - $47$) a pulse loop that maintains a memory that the decision was made. Logic populations: ``Logic - Trigger" ($49$) a population that is conditionally excited when both Trigger and Input are excited, and ``Conditional Output" ($50$) a population that is conditionally excited when both Hadamard coefficient $8$ and a population in Output Copy are excited. C) Raster plot showing spikes from the graded chain. $T_0 = 4$ ms, $T = 7.5$ ms, $\tau = 5$ ms. Time runs from left to right. D) Raster plot showing spikes from the gating chain.  E) Mean firing rates of the graded chain averaged over $50$ realizations. F) Mean firing rates for the gating chain averaged over $50$ realizations. E,F) The firing rates have been smoothed by a moving average process with width $2$ ms.}
\end{figure*}

To illustrate the capability of pulse-based information processing to perform complex computations, we show results from a neural circuit that, after being triggered by a streaming input, encodes and transforms the input then makes a decision based on the transformed input that affects subsequent processing (Fig. 5). The neural circuit consists of (see Fig. 5A) 1) a trigger, 2) a module used to keep sampled streaming input in short-term memory, 3) a $4 \times 4$ Hadamard transform (a Fourier transform using square-wave, Walsh functions as a basis), 4) a second set of Hadamard outputs (Hadamard Copy) representing output copy to a downstream circuit, 5) an Input population, 6) a Shut Down population to terminate processing, 7) a Compute gating chain to drive the computation, 8) a Vigilance gating chain that serves as a processing indicator and clock to synchronize the triggering of an output decision, 9) an Output Copy gating chain that serves as a decision indicator and is turned on based on the amplitude of the $8$'th Hadamard coefficient, and 9) Logic populations for triggering the computation and making the decision to copy the Hadamard output. Output then triggers circuit shutdown by inhibiting all gating chains.

In Fig. 5B, Memory designates Read In ($1$, $6$, $10$, $13$) and (non-cyclic) Memory populations. Hadamard designates populations holding Hadamard coefficient amplitudes. The Hadamard transform is divided into two parallel operations, one that results in positive coefficients, the other in absolute values of negative coefficients. Hadamard Copy designates populations into which the Hadamard transform may be copied. Input designates a population that linearly transduces a signal from outside the network. And Shutdown designates a population that receives summed input from the Hadamard Copy populations. Upon excitation, it shuts down the input and gating populations and terminates the computation.

Initially, the Trigger population is re-entrantly excited until the Input amplitude increases (as indicated in the top row of Fig. 5D). Input combines with Trigger to initiate firing in the Logic - Trigger population, which triggers the Compute gating chain and initiates the computation. Trigger is subsequently turned off by inhibition from the Compute gating chain. We show the computation for three successive windows, each of length $4T_0$. The gating chain binds the input into four memory chains of length $4T_0$, $3T_0$, $2T_0$ and $T_0$. Thus, four temporally sequential inputs are bound in four of the memory populations beginning at times $t = 4, 8, 12 T_0$. Hadamard transforms are performed beginning at $t = 5, 9, 13 T_0$. Each subsequent read in starts one packet length before the Hadamard transform so that the temporal windows are adjacent. At time $t = 0.06$ s, the high amplitude in Hadamard coefficient $8$ combines with gating population Conditional Output to initiate the Output Copy chain. The output is copied to Hadamard Copy populations, which then cause the shutdown of the gating chain.

This probabilistic, spike-based algorithm uses a self-exciting population coupled to a streaming input to trigger the computation (see Fig. 5B,C,D,E), then continuously gates $4$ sequential input amplitudes into $4$ read in populations and maintains the input values by gating them through working memory populations until all values are simultaneously in $4$ working memory populations. At this point, these values are gated to Hadamard populations transforming the input values into Hadamard coefficients (one set of positive coefficients and one set of absolute values of negative coefficients \cite{SornborgerWangTao}). At this point, a time-windowed Hadamard transform has been computed on the input. Gating pulses are interleaved such that this computation is performed iteratively on successive windows of length $4T$ from the streaming input.

To implement a conditional copy of the transformed data, we combine the output of the (arbitrarily chosen) eighth Hadamard coefficient in the present Hadamard output and the first population in the ``Vigilance" gating chain. This operation causes the graded pulse to activate the Output Copy chain when its amplitude is sufficiently high, conditionally causing a pulse to cascade through 4 gating populations with the last population gating the transfer from the subsequent Hadamard output to the 8 output neurons. Once the Hadamard output is copied, it activates the Shutdown population, which inhibits all populations in the gating chains, terminating the computation.

\section{Discussion}

The emerging picture from accumulating experimental evidence is that coherent activity is a fundamental contributor to cognitive function. In particular, coherence (alternatively, correlation of a signal at a given lag) is a measure of the efficacy of univariate information transfer between neuronal populations (a matrix-valued quantity is needed to measure the efficacy of multivariate information transfer). Here we have demonstrated a coherent transfer mechanism that dynamically routes graded information through a neural circuit using stereotyped gating pulses and performs computation via non-linear coupling of graded and gating pulses. As we have shown, SGSCs can be used as building blocks to implement complex information processing algorithms, including sub-circuits responsible for short-term memory, information processing and computational logic. As such, synfire-gated synfire chains should be considered as a candidate mechanism whenever coherent activity is implicated in information transfer.


Rapid visual categorization (RVC) experiments have demonstrated that objects can be recognized as early as $250 - 300$ ms after presentation. It has been conjectured that massively parallel, feedforward networks are used during RVC computations for maximum speed \citep{pmid11244543,pmid22007180,pmid25208739,pmid12684488}. At $40$ Hz, $10 - 12$ feedforward processing layers would be needed to construct such a network (Fig. 1). The signal-to-noise ratios that we demonstrated for the SGSC (Fig. 4) are good enough that it could be used for this type of information transfer at $40$ Hz. Indeed, in our examples, we show rapid propagation of graded information at $300$ Hz.

To our minds, the success of the SGSC graded information propagation mechanism rests on the structural robustness of the pulse gating mechanism. One contribution to robustness is that the synfire chain that is used for pulse generation approaches a fixed amplitude attractor with fixed temporal offset. A second contribution is that by providing overlapping temporal windows for information integration, the constraints on parametric precision to achieve graded information transfer are relaxed (Fig. 2C and related text). Having said that, the correspondence between our mean-field model and the SGSC gives weight to the idea that pulse-gating, independently of how it is implemented, is a robust mechanism for controlling information transfer in neural circuits. Thus, there is no particular reason that other pulse generators should not be entertained. For instance, experiments implicate the PVBC/OLM system of interneurons in cortical pulse generation \citep{pmid23010933}.

A conceptual framework for the manipulation of information in neural circuits arises naturally when one considers graded information transfer in the context of coherently interacting neuronal populations. In this framework, information processing and information control are conceived of as distinct components of neural circuits \citep{SornborgerWangTao}. This distinction has been used previously \cite{LismanIdiart1995,JensenLisman2005,pmid24730779,pmid25503492} in mechanisms for gating the propagation of fixed amplitude waveforms. Here, by providing a mechanism for the propagation of graded information and including computational logic by allowing graded and gating chains to interact, active linear maps (see Materials and Methods) take prominence as a key information processing structure.


It is worth mentioning that when we constructed the neural circuit example in Fig.\;5, we started at the algorithmic level, then implemented the algorithm in the mean-field firing rate model, then translated the mean-field model into the spiking, I\&F network. We feel that this is a major strength of the SGSC-based information processing framework, {\it i.e.} that it provides a practical pathway for designing computational neural circuits, either for the purpose of forming hypotheses about circuits in the brain, or for implementing algorithms on neuromorphic chips.

\section{Materials and Methods}

\subsection{The Synfire-Gated Synfire Chain Circuit}

Individual current-based, I\&F point neurons in the SGSC have membrane potentials described by
\begin{subequations}
\begin{equation}
\frac{d}{dt} v^\sigma_{i,j} = -g_{leak} \left( v^\sigma_{i,j} - V_{leak} \right) + \sum_{\sigma' = 1}^2 I^{\sigma \sigma'}_{i,j} + I^\sigma_{i,j}
\end{equation}
\begin{equation}
\tau \frac{d}{dt} I^{\sigma \sigma'}_{i,j} =  -I^{\sigma \sigma'}_{i,j} + \frac{S^{\sigma \sigma'}}{p_{\sigma \sigma'}N_{\sigma'}} \sum_{i'} \sum_l \delta \left( t - t^{\sigma',l}_{i',j-1} \right)
\end{equation}
\begin{equation}
\tau \frac{d}{dt} I^{\sigma}_{i,j} =  -I^{\sigma}_{i,j} + f^\sigma \sum_l \delta \left( t - s^{l}_{i,j} \right)
\end{equation}
\end{subequations}
where $\sigma, \sigma' = 1,2$ with $1$ for the graded chain and $2$ for the gating chain, $i = 1,\dots,N_\sigma$ denotes the number of neurons per population for each layer, $j = 1,\dots,M$ denotes the layer; individual spike times, $\{ t^{\sigma, l}_{i,j} \}$, with $l$ denoting spike number, are determined by the time when $v^{\sigma}_{i,j}$ reaches $V_{Thres}$. The parameters $g_{leak}$ and $V_{leak}$ denote the leak conductance and the leak potential. We have used reduced dimensional units in which time retains dimension in seconds and $V_{thresh} - V_{leak} = 1$. In these units $g_{leak} = 50$/sec. The parameter $\tau$ denotes the synaptic timescale ($\tau = 5$ ms, or approximately an AMPA synaptic timescale, in the Results above). The current $I^{\sigma \sigma'}_{i,j}$ is the synaptic current of the $\sigma$ population produced by spikes of the $\sigma'$ population. The parameter $S^{\sigma \sigma'}$ denotes the synaptic coupling strength and $p_{\sigma \sigma'}$ is the probability of coupling. $I_{i,j}^\sigma$ is a background noise current generated from Poisson spike times, $\{ s_{i,j}^l \}$, with strength $f^\sigma$ and rate $\nu_\sigma$.

\subsection{More Complex Synaptic Processing}

General SGSC circuits can incorporate a number of subcircuits, such as short-term memory and processing due to non-trivial synaptic connectivities \citep{SornborgerWangTao} such as the circuit shown in Fig. 5. In this case, more general connectivities are needed and the above equations become
\begin{subequations}
\begin{equation}
\frac{d}{dt} v^\sigma_{i,j} = -g_{leak} \left( v^\sigma_{i,j} - V_{leak} \right) + \sum_{\sigma' = 1}^2 I^{\sigma \sigma'}_{i,j} + I^\sigma_{i,j}
\end{equation}
\begin{equation}
\tau \frac{d}{dt} I^{\sigma \sigma'}_{i,j} =  -I^{\sigma \sigma'}_{i,j} + \frac{S^{1}}{p_{\sigma \sigma'}N_{\sigma'}} \sum_k K^{\sigma \sigma'}_{jk} \sum_{i'} \sum_l \delta \left( t - t^{\sigma',l}_{i',k} \right)
\end{equation}
\begin{equation}
\tau \frac{d}{dt} I^{\sigma}_{i,j} =  -I^{\sigma}_{i,j} + f^\sigma \sum_l \delta \left( t - s^{l}_{i,j} \right)
\end{equation}
\end{subequations}
Here, the synaptic connectivity for the graded chain is $K^{11}_{jk}$, the coupling between the chains is $K^{12}_{jk}$, and the connectivity of the gating chain is $K^{22}_{jk}$. Interaction between the graded chain and the gating chain is given by $K^{21}_{jk}$. We use $K^{21}_{jk}$ to implement conditional logic operations.

\subsection{Mean-field Solutions for Synaptic Current Propagation in the Overlapping Pulse Case}

To analyze graded propagation for the case in which the integration of graded information in successive populations overlaps in time, we assume that the gating pulse is square with amplitude sufficient to bring neuronal populations up to the firing threshold. We also assume that above threshold the activity function is linear \citep{SornborgerWangTao}. 

Firing in the upstream population is integrated by the downstream population. Thus, the downstream synaptic current obeys
$$\tau \frac{d}{{dt}}{I_d} =  - {I_d} + S{m_u } \; ,$$
where $S$ is a synaptic coupling strength, $I_d(t)$ is the downstream synaptic current, and $\tau$ is a synaptic timescale. In a thresholded-linear model, the upstream firing rate is
$$m_u \approx \left[ I_u(t) + I^{Exc}_0 - g_0 \right]^+ \; ,$$ 
where $I^{Exc}_0 = g_0 p(t)$ is an excitatory gating pulse, $p(t) = \theta(t) - \theta(t - T)$ and $\theta$ is the Heaviside step function, causing the downstream population to integrate $I_u(t)$, giving the current 
$$G \left[ I_d \right] (t) \equiv S e^{-t/\tau} \left[ \int_0^t ds \; e^{s/\tau} I_u(s) + c \right] \; .$$

The graded population is pulsed for time $T$. The offset between successive gating pulses is given by $T_0$ (see Fig. 2). In \citep{SornborgerWangTao}, we studied the case where $T = T_0$. That is, the downstream pulse turned on just when the upstream pulse turned off. Here, we focus on the case where $\eta = T/T_0 > 1$, and $\eta$ need not be an integer. Let $n$ be the integer part of $\eta$. Then $T = n T_0 + T_1$, where $T_1 < T_0$.

In the Appendices, we give a general derivation of time translationally invariant solutions in this context. In brief, a translationally invariant, graded current waveform is found for a particular feedforward strength, $S = S_{exact}$, by integrating the upstream firing rate over intervals of length $T_1, T_0, \dots, T_0$, while enforcing continuity of the solution. For these solutions, $S_{exact}$ is given by the smallest root of
\begin{equation*}
  \sum_{j=0}^n \frac{(-1)^j}{(n-j)!} \left[ \frac{\left((j+1) T_0 - T_1 \right)}{\tau} S e^{T_0/\tau} \right]^{n-j} = 0 \; .
\end{equation*}

\subsection{Information Processing Using Graded Transfer Mechanisms}

As we demonstrate in Results, current amplitude transfer for the SGSC may be modeled with a piecewise linear activity function, therefore synaptic connectivities between two layers each containing a vector of populations, perform a linear operation on the currents in the upstream layer \citep{SornborgerWangTao}. For instance, consider an upstream vector of neuronal populations with currents, $\mathbf{I}^{u}$, connected via a connectivity matrix $K$ to a downstream vector of neuronal populations, $\mathbf{I}^{d}$:
\begin{equation}
\mathbf{I}^{u}(t) \overset{K}{\rightarrow} \mathbf{I}^{d}(t) \; .
\end{equation}
With feedforward connectivity, $K$, the current amplitude, $\mathbf{I}^d$, obeys
\begin{equation}
\tau \frac{d}{dt} \mathbf{I}^{d} = -\mathbf{I}^{d} + S \left[ \sum_k K \mathbf{I}^u + \mathbf{p}^u(t) - g_0 \right]^+ \; ,
\end{equation}
where $\mathbf{p}^u(t)$ denotes a vector gating pulse on the upstream population. This results in the solution $\mathbf{I}^{d}(t-T) = P K \mathbf{I}^u(t)$, where $P$ is a diagonal matrix with the gating pulse vector, $\mathbf{p}$, of $0$s and $1$s on the diagonal indicating which neurons were pulsed during the transfer.

This discussion has identified three components of an information processing framework that naturally arises from mechanisms such as the SGSC: 
\begin{enumerate}[itemsep = -1mm]
\item information content - graded current, $\mathbf{I}$
\item information processing - synaptic weights, $K$
\item information control - pulses, $\mathbf{p}$
\end{enumerate}
Note that the pulsing control, $\mathbf{p}$, serves as a gating mechanism for routing neural information into (or out of) a processing circuit. We, therefore, refer to amplitude packets, $\mathbf{I}$, that are guided through a neural circuit by a set of stereotyped pulses as ``bound" information. In the SGSC, information content is carried by the graded chain ({\it e.g.} Fig. 5b,d), information processing is performed by the synaptic connectivities ({\it e.g.} Fig. 5a) and information control is performed by the gating chain ({\it e.g.} Fig. 5c,e). We will refer to the combination of these control and processing structures as {\it active linear maps}.

In order to make a decision, non-linear logic circuits are introduced. A simple decision can be implemented in our framework by allowing interaction between information control and content. In our example, a graded and a gating pulse were combined to make a decision, then the output was fed as input to a gating chain. If the graded chain output a low value, the gating chain was not switched on. However, if the graded chain output was high, this initiated pulses in the gating chain, which rapidly approached an attractor. Thus, the interaction caused conditional firing in the gating chain.

\begin{acknowledgments}
{\bf Acknowledgements}  L.T. thanks the UC Davis Mathematics Department for its hospitality. A.T.S. would like to thank Liping Wei and the Center for Bioinformatics at the College of Life Sciences at Peking University
for their hospitality. This work was supported by the Ministry of Science and Technology of China through the Basic Research Program (973) 2011CB809105 (W.Z. and L.T.), by the Natural Science Foundation of China grant 91232715 (W.Z. and L.T.) and by the National Institutes of Health, CRCNS program NS090645 (A.T.S. and L.T.).
\end{acknowledgments}

\bibliographystyle{apalike}
\bibliography{Biblio}

\section*{Appendices}

\subsection*{Appendix 1: Graded Propagation via Overlapping Gating Pulses}

\subsubsection*{Mean-Field Analysis}
To analyze graded propagation for the case in which the integration of graded information in successive populations overlaps, we assume that the gating pulse is square with amplitude sufficient to bring neuronal populations up to the firing threshold. We also assume that above threshold the activity function is linear \cite{SornborgerWangTao}.

Firing in the upstream population is integrated by the downstream population. The downstream synaptic current obeys
$$\tau \frac{d}{{dt}}{I_d} =  - {I_d} + S{m_u } \; ,$$
where $S$ is a synaptic coupling strength, $I_d(t)$ is the downstream synaptic current, and $\tau$ is a synaptic timescale.  For simplicity, we set $\tau = 1$. Therefore, from here on, time will be measured in units of $\tau$. The upstream firing rate is
$$m_u \approx \left[ I_u(t) + I^{Exc}_0 - g_0 \right]^+ \; ,$$ 
where $I^{Exc}_0 = g_0 p(t)$ is an excitatory gating pulse, $p(t) = \theta(t) - \theta(t - T)$ and $\theta$ is the Heaviside step function, causing the downstream population to integrate $I_u(t)$, giving the current 
$$G \left[ I_d \right] (t) \equiv S e^{-t} \left[ \int_0^t ds \; e^s I_u(s) + c \right] \; .$$ 
As shown in Fig. 2 (in the main text), we label the offset between upstream and downstream gates by $T_0$. The solution is broken into a set of $n+1$ equal intervals of length $T_0$ labelled from later to earlier in time. The $n+1$'st (earliest) interval typically extends before the onset of the pulse, since an arbitrary pulse's length is not an integer multiple of the lag $T_0$. The delay between the beginning of the final interval and the beginning of the pulse is denoted $T_1$ (see Fig. 2b). During each interval, the downstream current integrates upstream firing shifted by $T_0$. The integrated currents and matching conditions for each interval solution are given by
\begin{eqnarray}
   G\left[ I_0 \right](t) =& I_1(t), & I_0(0) = I_1(T_0) \nonumber \\
   G\left[ I_1 \right](t) =& I_2(t), & I_1(0) = I_2(T_0) \nonumber \\
     \vdots && \nonumber \\
   G\left[ I_{n-1} \right](t) =& I_{n}(t), & I_{n-1}(0) = I_{n}(T_0) \nonumber \\
   G\left[ I_{n}(s) \theta(s-T_1) \right](t) =& I_{n+1}(t), & I_{n}(0) = I_{n+1}(T_0) \nonumber \\
   && I_{n+1}(T_1) = 0 \; .\nonumber
\end{eqnarray}
Here, due to our definition of $G$, each interval solution begins at time $t = 0$. This leads to the particular form of the matching conditions above. The interval solution will be shifted later on to give a continuous function, made up of the integral solutions, for the synaptic current.

Evaluating, we find
\begin{eqnarray*}
I_0(t) &=& S e^{-t} c_0 \\
I_1(t) &=& S e^{-t} \left[ \int_0^t ds \; e^s e^{-s} S c_0 + c_1 \right] \\
          &=& S e^{-t} \left[ S c_0 t + c_1 \right] \\
&\vdots& \\
I_j(t) &=& S e^{-t} \sum_{i=0}^j c_i \frac{(St)^{j-i}}{(j-i)!} \\
&\vdots& \\
I_{n}(t) &=& S e^{-t} \sum_{i=0}^{n} c_i \frac{(St)^{n-i}}{(n-i)!} \\
I_{n+1}(t) &=& S e^{-t} \left[ \int_{T_1}^t ds \; e^s S e^{-s} \sum_{i=0}^{n} c_i \frac{(Ss)^{n-i}}{(n-i)!} \right]\\ &=& S e^{-t} \sum_{i=0}^{n} c_i \frac{S^{n-i+1}}{(n-i+1)!}(t^{n-i+1} - T_1^{n-i+1}) \; ,
\end{eqnarray*}
where $I_0(t)$ is a current due to the exponential decay of the upstream population's firing rate after its integration. Applying the matching conditions, we obtain equations for the coefficients, $c_i$,
\begin{equation*}
I_0(0) = I_1(T_0) \Rightarrow c_0 = e^{-T_0} \left[ S c_0 T_0 + c_1 \right]
\end{equation*}
and, in general,
\begin{equation*}
I_{j-1}(0) = I_j(T_0) \Rightarrow c_{j-1} = e^{-T_0} \sum_{i=0}^j c_i \frac{(ST_0)^{j-i}}{(j-i)!} \; .
\end{equation*}
Matching conditions on the final segment
\begin{eqnarray*}
I_{n}(0) &=& I_{n+1}(T_0) \\
&\Rightarrow& c_{n} = e^{-T_0} \sum_{i=0}^{n} c_i \frac{S^{n-i+1}}{{n-i+1}!} (T_0^{n-i+1} - T_1^{n-i+1}) \\
I_{n+1}(T_1) &=& 0 \Rightarrow c_{n+1} = 0 \\
\end{eqnarray*}
Translating to matrix notation, we define
\begin{widetext}
\begin{equation*}
M = \left[ \begin{array}{ccccccc}
           e^{-T_0} S T_0 - 1 & e^{-T_0} & 0 & \dots & \dots & \dots & 0 \\
           e^{-T_0} S^2 \frac{T_0^2}{2} & e^{-T_0} S T_0 - 1 & e^{-T_0} & 0 & \dots & \dots & 0 \\
           e^{-T_0} S^3 \frac{T_0^3}{6} & e^{-T_0} S^2 \frac{T_0^2}{2} & e^{-T_0} S T_0 - 1 & e^{-T_0} & 0 & \dots & 0 \\
           \vdots & \ddots & \ddots & \ddots & \ddots & \ddots & \vdots \\
           e^{-T_0} S^{n} \frac{T_0^{n}}{(n)!} & e^{-T_0} S^{n-1} \frac{T_0^{n-1}}{(n-1)!} & \dots & \dots & \dots & \dots & e^{-T_0}\\
           e^{-T_0} S^{n+1} \frac{(T_0^{n+1} - T_1^{n+1})}{(n+1)!} & e^{-T_0} S^{n} \frac{(T_0^{n} - T_1^{n})}{(n)!} & \dots & \dots & \dots & \dots & e^{-T_0} S (T_0 - T_1) - 1
       \end{array} \right]
\end{equation*}
\end{widetext}
and
$$\mathbf{c} = [c_0 \; c_1 \; \dots \; c_{n}]^T \; .$$

We then solve $\det (M) = 0$ to determine the values, $S_i$, for which a solution exists. In general, the determinantal equation is an $n + 1$'st order polynomial in $S$. Solution vectors may be found by solving $$M(S_i) \mathbf{c} = \mathbf{0} \; .$$ For  $n < 16$ (as far as we have checked), we find that all $\{ S_i \}$ are positive and real and only the smallest eigenvalue, $S_{min}$, corresponds with a solution vector of all positive coefficients, $c_i$. Thus, we conjecture that there is just one graded solution for a given $T$.

\subsection*{General Expression for Determinant of $M$}
The matrix $M$ is a full Hessenberg matrix with one upper diagonal and all filled lower triangular elements. A general recursive expression for the determinant may be derived in the standard way, i.e. to use row operations (of determinant $1$) to reduce $M$ to diagonal form. Then, the determinant is the product of the values, $\{ \lambda_i \}$, along the diagonal, $\prod_i \lambda_i$.

We will work with the matrix $M' \equiv e^{-T_0} M$. Note that $\det{M} = e^{n T_0} \det{M'}$. With $a_1 = ST_0 - e^{T_0}$, $a_j = \frac{(ST_0)^j}{j!}, j > 1$ and $b_j = \frac{(ST_1)^j}{j!}, j \ge 1$, we have
\begin{equation*}
M' = \left[ \begin{array}{ccccccc}
            a_1 & 1 & 0 & \dots & \dots & \dots & 0 \\
            a_2 &  a_1 & 1 & 0 & \dots & \dots & 0 \\
            a_3 &  a_2 &  a_1 & 1 & 0 & \dots & 0 \\
           \vdots & \ddots & \ddots & \ddots & \ddots & \ddots & \vdots \\
            a_n &  a_{n-1} & \dots & \dots & \dots & a_1 & 1 \\
            a_{n+1} - b_{n+1} &  a_n - b_n & \dots & \dots & \dots & \dots &  a_1-b_1
       \end{array} \right]
\end{equation*}
The super diagonal $1$s are now eliminated row by row, giving diagonal terms 
\begin{eqnarray*}
\lambda_1 & = & a_1 - b_1 \\
\lambda_2 & = & a_1 - \frac{a_2 - b_2}{\lambda_1} \\
\lambda_3 & = & a_1 - \frac{a_2 - \frac{a_3 - b_3}{\lambda_1}}{\lambda_2} \\
& \vdots &
\end{eqnarray*}
After dividing through by a common denominator, we find
\begin{equation*}
   \lambda_j = \prod_{m=1}^{j-1} \frac{1}{\lambda_m} \left[ \sum_{k=1}^j (-1)^{k-1} a_k \prod_{p=1}^{j-k} \lambda_p + (-1)^j b_j \right]
\end{equation*}
This gives a recursion relation for the determinant, $d_n \equiv \det{M'_n}$, where $M'_n$ denotes the $n \times n$ lower-right hand submatrix of $M'$,
\begin{equation*}
d_n \equiv \prod_{m=1}^n \lambda_m = \left[ \sum_{k=1}^n (-1)^{k-1} a_k d_{j-k} + (-1)^n b_n \right]
\end{equation*}
From an investigation of the low order values of this expression, we conjecture that
\begin{equation*}
  d_n = \sum_{j=0}^n \frac{(-1)^j}{(n-j)!} \left[ \left((j+1) T_0 - T_1 \right) S e^{T_0} \right]^{n-j} \; ,
\end{equation*}
which we have checked for $n < 16$.

\subsection*{Appendix 2: Network Parameters for Neural Circuit in Figure 5}

Below we provide the network parameters for the simulation presented in Fig. 5.

Population sizes are: $N_1 = 1000$, $N_2 = 100$

Coupling strengths and probabilities are $S^{11} = 2$ with $K^{11}_{jk} = 1$ for memory and Hadamard Copy populations. $K^{11}_{jk} = 0.2$ for the first layer, and $K^{11}_{jk} = \pm 0.5$ for other layers in Hadamard population. $K^{11}_{jk} = 1.3$ for the connectivity from the Hadamard Copy to Shut down population.

The coupling probability $p_{11} = 0.01$.

$S^{22} = 2.7$ with $K^{22}_{jk} = 1$ for all gating populations (Compute, Vigilance and Output Copy). $p_{22} = 0.8$. There is a time delay between layers in the gating population with $t_{delay} = 1$ms. For the Trigger population, we used the same parameters as in the gating populations ($N = 100$, $S = 2.7$, $p = 0.8$, $\nu = 25$Hz, $f = 0.2$, $t_{delay} = 0$, no self-inhibition). The Trigger population is inhibited by the first gating population with $S^{22} = 2.7$, $K^{22}_{jk} = -3.7$.
$S^{12} = 0.37$, with $K^{12}_{jk} = 1$ for the connectivity between gating chain and graded chain. $p_{12} = 0.01$. 

For the logic populations, $S^{21} = 2$, $K^{21}_{jk} = 1$ and $p_{21} = 1$. The Logic - Trigger population
receives inputs from both the Trigger and the Input. The Logic - Conditional Output population receives inputs from the $8$th Hadamard population and the first Vigilance population. For the Shutdown population, $S^{21} = 3.4$, $K^{21}_{jk} = -2$, and $p_{21} = 1$. Note that the shut down population inhibits all gating populations and the Input population.

There is self-inhibition for all neurons with $S^{11} = 2$, $K^{11}_{jk} = -0.6$; $S^{22} = 2.7$, $K^{22}_{jk} = -0.5$.

The background Poisson inputs are $\nu^1 = 25$Hz, $f^1 = 0.2$; $\nu^2 = 25$Hz, $f^2 = 0.2$. 
For all neurons, the refractory period is 2ms.

\end{document}